\documentclass[a4paper, 11pt, oneside]{article}
\pdfoutput=1
\usepackage{jheppub}		
\usepackage{graphicx,wrapfig,float,slashed,indentfirst}
\usepackage{amsmath}
\usepackage{xcolor,bbm}
\usepackage{fancyhdr}	
\usepackage{epstopdf}
\usepackage{xr}
\usepackage{graphics}
\usepackage[mathscr]{eucal}
\usepackage{xspace}
\usepackage{listings}
\usepackage{gensymb}
\usepackage{pstricks}
\usepackage{soul}
\usepackage{xspace}
\usepackage{braket}
\usepackage{slashed}
\usepackage{subfigure}
\usepackage{array}
\usepackage{feynmp-auto}
\usepackage[makeroom]{cancel}
\newcommand{\nc}{\newcommand}

\nc{\beq}{\begin{equation}}  \nc{\eeq}{\end{equation}}
\nc{\bea}{\begin{eqnarray}}  \nc{\eea}{\end{eqnarray}}
\nc{\baa}{\begin{array}}     \nc{\eaa}{\end{array}}
\nc{\bit}{\begin{itemize}}   \nc{\eit}{\end{itemize}}
\nc{\ben}{\begin{enumerate}} \nc{\een}{\end{enumerate}}
\nc{\bce}{\begin{center}}    \nc{\ece}{\end{center}}
\nc{\bpm}{\begin{pmatrix}}   \nc{\epm}{\end{pmatrix}}
\nc{\bvt}{\begin{verbatim}}  \nc{\evt}{\end{verbatim}}
\nc{\non}{\nonumber} 
\newcolumntype{M}{>{$\vcenter\bgroup\hbox\bgroup}c<{\egroup\egroup$}}

\def\mev{\;\hbox{MeV}}
\def\gev{\;\hbox{GeV}}

\def\cm{\;\hbox{cm}}
\def\s{\;\hbox{s}}

\def\km{\;\hbox{km}}

\def\diag{\hbox{\diag}}
\def\lsp{\;\;\;\;}

\def\zBB{{\mathbb{Z}}}

\def\z2{\zBB_2}

\def\mpl{M_{Pl}}
\def\mone{M_{h_1}}
\def\mtwo{M_{h_2}}
\def\zp{Z^\prime}
\def\mzp{M_{\zp}}
\def\vx{v_x}
\def\gx{g_x}
\def\mdm{M_{DM}}

\def\msm{125.7}

\def\odm{\Omega h^2}
\def\uone{U(1)_X}

\def\vr{v_{\rm rel}}
\def\sigv{\langle\sigma\vr\rangle}
\def\sigvp{\langle\sigma\vr\rangle_0}
\def\tkd{T_{kd}}
\def\tdm{T_{DM}}

\def\xkd{x_{kd}}

\def\ss{\sigma_{\rm self}/\mzp}
\def\sdm{\sigma_{\rm self}/M_{DM}}
\def\sigs{\sigma_{\rm self}}
\def\vr{v_{\rm rel}}

\def\lsim{\mathrel{\raise.3ex\hbox{$<$\kern-.75em\lower1ex\hbox{$\sim$}}}}
\def\gsim{\mathrel{\raise.3ex\hbox{$>$\kern-.75em\lower1ex\hbox{$\sim$}}}}

\def\ot#1{%
  \mathrel{\vbox{\offinterlineskip\ialign{%
    \hfil##\hfil\cr
    $\scriptscriptstyle(\,\sim\,)$\cr
    \noalign{\kern-.1ex}
    $#1$\cr
}}}}

\def\inv#1{\frac1{#1}}
\title{Resonance enhancement of dark matter interactions:\\
the case for early kinetic decoupling and velocity dependent resonance width}
\author{M. Duch and}
\author{B. Grzadkowski}
\affiliation{Faculty of Physics, University of Warsaw, Pasteura 5, 02-093 Warsaw, Poland}
\emailAdd{mateusz.duch@fuw.edu.pl}
\emailAdd{bohdan.grzadkowski@fuw.edu.pl}
\date{\today}

\abstract{
Motivated by the possibility of enhancing dark matter (DM) self-interaction cross-section $\sigma_{\rm self}$, we have revisited the issue of DM annihilation through a Breit-Wigner resonance. In this case thermally averaged annihilation cross-section has strong temperature dependence, whereas elastic scattering of DM on the thermal bath particles is suppressed. This leads to the early kinetic decoupling of DM and an interesting interplay in the evolution of DM density and temperature that can be described by a set of coupled Boltzmann equations. The standard Breit-Wigner parametrization of a resonance propagator is also corrected by including momentum dependence of the resonance width. It has been shown that this effects may change predictions of DM relic density by more than order of magnitude in some regions of the parameter space.
Model independent discussion is illustrated within a theory of Abelian vector dark matter. The model assumes extra $U(1)$ symmetry group factor and an additional complex Higgs field needed to generate a mass for the dark vector boson, which provides an extra neutral Higgs boson $h_2$.  
We discuss the resonance amplification of $\sigma_{\rm self}$. It turns out that if DM abundance is properly reproduced, the Fermi-LAT data favor heavy DM and constraint the enhancement of $\sigma_{\rm self}$ to the range, which cannot provide a solution to the small-scale structure problems.}


\begin{document}

\maketitle

\section{Introduction}

The dark matter (DM) constitutes 84.5\% of total Universe matter~\cite{Ade:2015xua}, nevertheless 
its origin is still unknown in spite of unprecedented effort made both by experimentalists and theoreticians. 
Even more, the standard WIMP (weakly interacting massive particles)  paradigm seems to suffer from various difficulties when confronted with observations on small cosmological scales.
For instance ``too-big-to-fail'' \cite{BoylanKolchin:2011de,Garrison-Kimmel:2014vqa} and the ``cusp-core'' \cite{Moore:1994yx,Flores:1994gz,Oh:2010mc,Walker:2011zu} problems are widely discussed in  the literature.  In particular, the DM densities inferred in central regions of DM dominated galaxies are usually smaller than expected from WIMP simulations \cite{Rocha:2012jg,Weinberg:2013aya}. 
It turns out that an appealing alternative is to assume that dark matter may self-interact strongly \cite{Spergel:1999mh}. The assumption of self-interacting dark matter (SIDM) implies that central (largest) DM density could be reduced.
The numerical simulations have shown that SIDM halos are consistent with observations if DM particles have a nuclear-scale self-interaction cross-section $0.1\cm^2/{\rm g}<\sdm<10\cm^2/{\rm g}$ within halos \cite{Markevitch:2003at,Rocha:2012jg,Vogelsberger:2012ku,Peter:2012jh,Zavala:2012us,Vogelsberger:2014pda,Buckley:2014hja,Elbert:2014bma,Harvey:2015hha,Kahlhoefer:2015vua}.
The largest values of $\sdm$ are in contradiction with the cluster limit, $\sdm <1.25\;\text{cm}^2\text{g}^{-1}$ \cite{Randall:2007ph}, therefore scenarios with velocity-dependent $\sdm$ are preffered \cite{Kaplinghat:2015aga}.

The aforementioned problems directly encouraged us to study the possibility for the Breit-Wigner resonant enhancement of dark matter annihilation and self-scattering\cite{Bento:2001yk,MarchRussell:2008tu,Ibe:2008ye,Ibe:2009dx,Ibe:2009en,Guo:2009aj,Bi:2009uj,Backovic:2009rw,Braaten:2013tza,Campbell:2015fra,Choi:2017mkk}. 
However, it turns out that physics relevant in the vicinity of a resonance is interesting on its own and worth studying in details regardless of any phenomenological applications.  
Particularly intriguing consequence of the resonance annihilation is the early kinetic decoupling of dark matter, which requires to examine the connection between DM density and temperature evolution. 
Even though the issues mentioned above are not 	limited to any particular model, and should be verified in each given case, we have chosen to illustrate them within a model of Abelian vector dark matter.

The paper is organized as follows.
In sec.~\ref{res_ann} we review the resonant annihilation process including velocity-dependent corrections to the standard 
Breit-Wigner description of a resonance. In sec.~\ref{kin_dec} the early kinetic decoupling of dark matter is discussed. 
In sec.~\ref{resonance} we consider the possibility of enhancement of dark-matter self-scattering cross section by a resonant 
s-channel Higgs boson exchange. Sec.~\ref{sig_self_fin} contains discussion of constraints relevant in the parameter space of DM with resonant annihilation.
Finally in sec.~\ref{con} we summarize our results.

\section{Resonant annihilation of DM}
\label{res_ann}

It is well known that the case of dark matter annihilation in the vicinity of a resonance requires a special treatment. In particular the standard velocity expansion of the cross-section fails and cannot be used to compute approximate thermal average in the relevant temperature range \cite{Griest:1990kh,Gondolo:1990dk}. Furthermore, the density of dark matter in the comoving volume can decrease substantially even after dark matter chemical decoupling, as thermally averaged cross-section increases with falling temperature. Finally the current annihilation rates can be larger by many orders of magnitude from the typical value $\sigvp = 3\times 10^{-26} \cm^3 \s^{-1}$ in the non-resonant case. This boost effect has been discussed, as an explanation to anomalies reported by cosmic rays experiments \cite{Pospelov:2008jd,Feldman:2008xs,Ibe:2008ye,Ibe:2009dx,Ibe:2009en,Guo:2009aj,Bi:2009uj}.

In this work we focus on two effects, which 
turn out to be important for DM annihilation especially if enhanced DM self-interaction is desirable.
The first one is early kinetic decoupling, which comes from the suppressed coupling between dark matter and visible sector in the highly resonant scenario and results in reducing DM temperature with respect to the SM thermal bath. The second effect is the energy-dependence of the resonance width, which cannot be neglected when the width is dominated by the decay into DM states. This scenario arises, when we keep couplings in the hidden sector unsuppressed what enforces the small coupling of the resonance to the SM particles.
\begin{center}
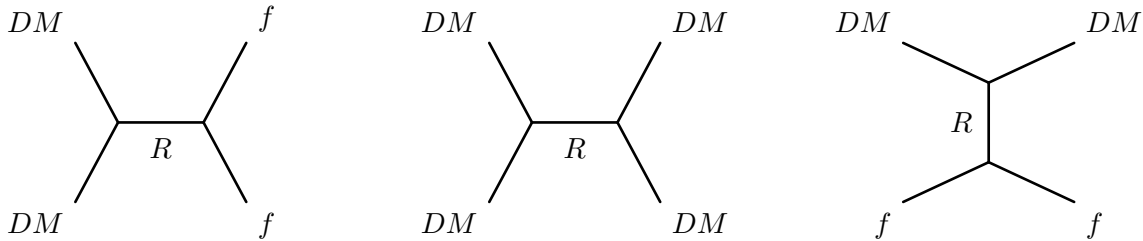
\begin{figure}[h!]
\vspace{.1cm}
\centering
\begin{tabular}{ccc}
\centering
\begin{fmffile}{hh2}
	        \begin{fmfgraph*}(80,60)
	            \fmfleft{i,j}
	            \fmfright{k,l}
	            \fmf{plain}{i,v1}
		    \fmf{plain}{j,v1}
		    \fmf{plain,label=$R$}{v1,v2}
		    \fmf{plain}{k,v2}
		    \fmf{plain}{v2,l}
		    \fmflabel{$DM$}{i} 
	        \fmflabel{$DM$}{j} 
	        \fmflabel{$f$}{k}
	        \fmflabel{$f$}{l} 
	        \end{fmfgraph*} 
\end{fmffile} \hspace{1.7cm} & 
	   
\begin{fmffile}{hh3}
	        \begin{fmfgraph*}(80,60)
	            \fmfleft{i,j}
	            \fmfright{k,l}
	            \fmf{plain}{i,v1}
		    \fmf{plain}{j,v1}
		    \fmf{plain,label=$R$}{v1,v2}
		    \fmf{plain}{k,v2}
		    \fmf{plain}{v2,l}
		    \fmflabel{$DM$}{i} 
	        \fmflabel{$DM$}{j} 
	        \fmflabel{$DM$}{k}
	        \fmflabel{$DM$}{l} 
	        \end{fmfgraph*} 
\end{fmffile} \hspace{1.7cm} & 	   

\begin{fmffile}{hh4}
	        \begin{fmfgraph*}(80,60)
	            \fmfleft{i,j}
	            \fmfright{k,l}
	            \fmf{plain}{j,v2}
		    \fmf{plain}{i,v1}
		    \fmf{plain,label=$R$}{v2,v1}
		    \fmf{plain}{l,v2}
		    \fmf{plain}{v1,k}
		    \fmflabel{$f$}{i} 
	        \fmflabel{$DM$}{j} 
	        \fmflabel{$f$}{k}
	        \fmflabel{$DM$}{l} 
	        \end{fmfgraph*} 
\end{fmffile} \hspace{1cm} 
\end{tabular}
\vspace{0.4cm}
\caption{\label{feyngraphs}
Feynman diagrams mediated by the resonance, which are responsible for DM annihilation, self-interaction and maintaining the kinetic equilibrium.}	
\end{figure}
\end{center}
\vspace{-1cm}
We will present detailed analysis of the dark matter evolution, when its annihilation is resonantly enhanced by a Breit-Wigner (BW) resonance in the s-channel. We limit ourselves to interactions, which are in the first approximation velocity-independent beyond the resonance (s-wave annihilation), i.e. the only velocity-dependence will come from the resonance propagator. 

In the Breit-Wigner approximation, the cross-section for a $2\to 2$ s-channel annihilation (fig. \ref{feyngraphs}) at a resonance of mass $M$ with identical dark matter particles of mass $m_i$ in the initial state and $m_f$ in the final state, can be written as (we follow the notation of \cite{Ibe:2008ye}\footnote{Extra factor of 2 comes from the assumption of identical particles in the initial state.}):
\begin{equation}
\sigma = \sum_{f\neq i}\frac{32\pi\omega\beta_f}{s\beta_i\bar{\beta}_i \bar{\beta}_f}\frac{M^2 \Gamma^2 B_i B_f}{(s-M^2)^2+\Gamma^2 M^2}
\label{csann}
\end{equation}
where $\Gamma$ is the total decay width of the resonance, $B_{i,\,f}$ are branching ratios for resonance decaying into initial and 
final states respectively, $\beta_{i,\,f}\equiv \sqrt{1-4 m_{i,\, f}^2/s}$, $\bar{\beta}\equiv\beta|_{s=M^2}$ and the statistical factor $\omega=(2J+1)/(2S+1)^2$ 
depends on spin of the resonance $J$ and spin of colliding particles $S$. 


If $2m_{i}>M$, then the decay of the resonance into initial states is not kinematically allowed. One can then treat $B_i$ and $\bar{\beta}_i$ as analytical continuations from the physical region $2m_{i}<M$ or one could express the cross-section in terms of the coupling between the initial state and the resonance mediator. In order to facilitate further discussion we will define the following dimensionless quantities
\beq
\eta_{i/f}\equiv\frac{\Gamma B_{i/f}}{M\bar{\beta}_{i/f}},\;\;\;\; \delta\equiv \frac{4m_i^2}{M^2} - 1\;\;\;\;\;\text{and}\;\;\;\;\gamma=\frac{\Gamma}{M},
\label{eta}
\eeq
where $\eta_{i/f}$ parametrizes the coupling between the initial (final) state and $\delta$ describes the position of the resonance pole. 


We are mostly interested in those regions of the phase space where DM particles are non-relativistic, 
so when $\vr \ll 1$. Large velocities may only appear in calculations of thermally averaged cross-sections, but there their contribution is negligible since it is very efficiently suppressed by the Maxwell-Boltzmann distribution. Therefore, hereafter we are going to adopt non-relativistic approximations. Since our work has been motivated by possibility of resonance enhancement of DM self-interactions therefore we will also focus on the vicinity of a resonance, i.e. $\delta \ll 1$.
In the non-relativistic approximation we obtain:
\begin{equation}
s=4m_i^2+m_i^2 \vr^2,\;{\rm where}\;\;\;\vr= 2\sqrt{\frac{s}{4 m^2_i}-1}\approx 2\beta_i.
\end{equation}
To first order in small parameters ($|\delta|$, $\vr\ll 1$) we obtain the following formula, which will be used in further discussions
\begin{equation}
\sigma\vr=\sum_{f\neq i}\frac{64\pi\omega}{M^2}\frac{\eta_i\eta_f\beta_f}{(\delta+\vr^2/4)^2+\gamma^2}.
\label{sigmav}
\end{equation}

One has to keep in mind that the resonant cross-section in a Breit-Wigner form is an approximation. The full propagator is obtained by the resummation of an infinite series of 1PI self-energy graphs $\Sigma(s)$ and contains an energy-dependent self-energy,
which in the BW propagator is replaced by the constant total decay width
\beq
\Gamma\equiv\Gamma(s=M^2),\;\;{\rm where}\;\; \Gamma(s)\equiv\frac{\Im{\Sigma(s)}}{M}.
\label{gams}
\eeq
This approximation is justified, when all thresholds for the resonance decay are far away from the pole position, $s=M^2$. We assume this is the case for the annihilation products, but the threshold for the decay into DM is always nearby for $|\delta|\ll 1$, therefore one has to include the leading energy dependence of $\Gamma(s)$.
By the optical theorem, each of the decay channels contributes to $\Gamma(s)$ with a phase space $\beta_{i,f}$ multiplied by a factor, which we approximate using constant $\eta_{i,f}$\footnote{This is justified by our assumption of a dominant s-wave annihilation beyond the resonance.}. We define energy-dependent width as
\beq
\gamma(s)\equiv\frac{\Gamma(s)}{M}\approx \eta_i\frac{\vr}{2}\theta(s-4m_i^2) + \sum_{f\neq i}\eta_f\bar\beta_f,
\label{swidth}
\eeq
where we have used $\beta_i=\vr/2$ and approximated contribution to $\Gamma(s)$ from the annihilation products by constants.
This gives the non-relativistic formula, where we neglect corrections of order $\delta$ and $v$ to $\eta_i$ and $\eta_f$. 

The extra difficulty comes from the gauge-dependence of vector contributions to the self-energy in theories with spontaneous symmetry breaking (as exemplified later by the vector dark matter model used to illustrate results of this paper). To get the modified self-energy, which fulfills the required physical properties, one has to cast the relevant amplitudes via the pinch technique algorithm (PT) \cite{Papavassiliou:1995fq,Papavassiliou:1997fn,Papavassiliou:1997pb}. However, as it has to display only the physical thresholds and coincide with gauge-independent value at the pole ($s=M^2$), the PT self-energy can also be approximated by (\ref{swidth}). 
We have checked by numerical calculations that the relativistic formula for thermally averaged cross-section $\sigv$  calculated in the $R_\xi$ gauge which does not contain unphysical thresholds in the resonance region ($\xi>1$), leads to the relic density that is within $10\%$ of the one obtained using non-relativistic approximation.

The energy-dependent width (\ref{swidth}) differs from the standard decay width only by a velocity $\vr$ which replaces constant factor $2 \bar\beta_i=2\sqrt{-\delta}$ that is present for negative~$\delta$. Nevertheless, as the low-energy behavior is crucial in the DM annihlation, we will show that this correction is essential, if decay of a resonance into DM dominates the decay width.

The cross-section (\ref{sigmav}) with $\gamma$ replaced by (\ref{swidth}) needs to be thermally averaged with the Maxwell-Boltzmann distribution. In the CM frame adopting consequently the non-relativistic approximation the average reads
\begin{equation}
\sigv_x = \frac{x^{3/2}}{2\sqrt{\pi}}\int_{0}^{\infty} d\vr \vr^2 e^{-x \vr^2/4} \sigma \vr.
\label{ther_aver}
\end{equation}

\begin{figure}[tb]
\centering
\includegraphics[width=0.49\textwidth]{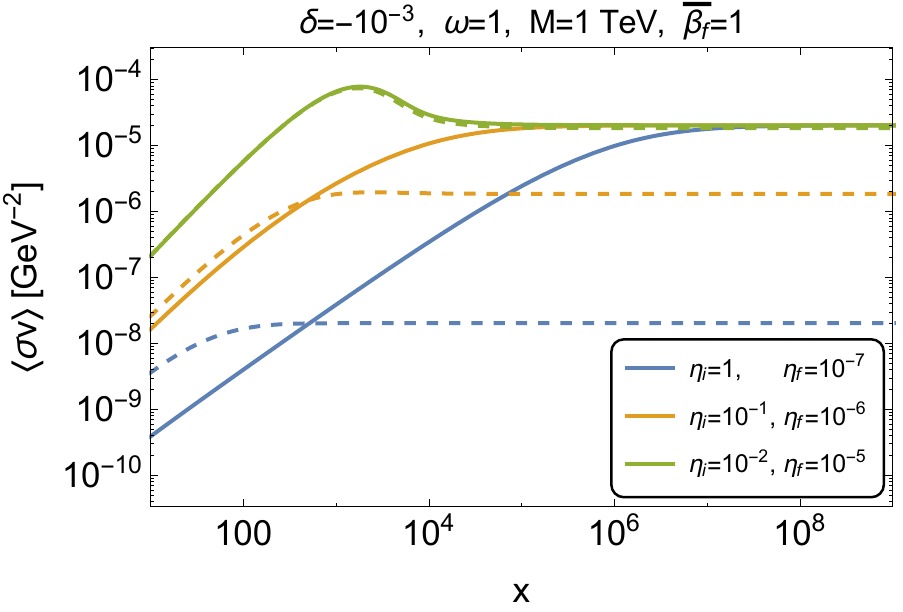}
\includegraphics[width=0.49\textwidth]{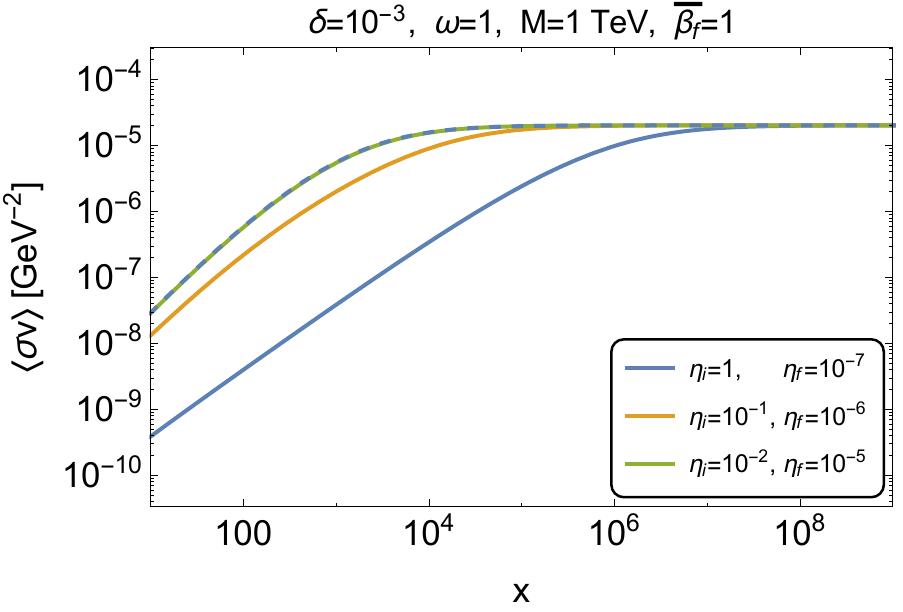}
\caption{\label{sigvplot}Thermally averaged annihilation cross-section $\sigv(x)$ for negative (left panel) and positive (right panel) value of $\delta$. The solid lines were obtained using the resonance formula (\ref{sigmav}) with energy-dependent width (\ref{swidth}) and dashed lines refer to constant width approximation. In the right panel all dashed lines coincide.}
\end{figure}

In fig.~\ref{sigvplot} we illustrate $\sigv(x)$ for selected values of $\delta$ and $\eta_{i/f}$ assuming annihilation into one effectively massless product ($\beta_f=\bar\beta_f=1$). To explain the qualitative behavior we examine the velocity dependence of the denominator in (\ref{sigmav})

\beq
\left(\delta+\frac{\vr^2}{4}\right)^2 + [\gamma(\vr)]^2 = \delta^2 + \eta_f^2 \bar\beta^2_f +\eta_i\eta_f\bar\beta_f \vr+ \left(\frac{\delta}{2}+\frac{\eta^2_i}{4}\right)\vr^2 + \frac{\vr^4}{16}. 
\label{deno} 
\eeq
Correct DM relic density in the scenario with resonant annihilation requires small DM-SM coupling, therefore we focus on the region where $\eta_i\eta_f\bar\beta_f\ll \delta^2$, so the linear term in velocity can be neglected in (\ref{deno}). 
Expanding the cross-section in powers of $\vr^2$ and keeping in mind that $\vr^2$ in the cross-section corresponds approximately to $1/x$ in the thermal average, one finds that $\sigv$ grows proportionally to $x$, before it reaches its low temperature limit around $x\gtrsim|\delta+\eta_i^2/2|/[2(\delta^2+\eta_f^2\bar\beta_f^2)]$. 
If $\delta<0$ and $\eta^2_i<2|\delta|$ then additionally a maximum appears at this value. We note that if $|\delta|\ll \eta^2_i/2$ the result does not depend on the sign of $\delta$.
The growth of the averaged cross-section leads to the extended period of DM annihilation and has impact on the bounds coming from the indirect searches. As we will discuss in the next section, it leads also to non-trivial temperature evolution of the hidden sector.

\section{Dark matter evolution and early kinetic decoupling}
\label{kin_dec}

In this work we assume that dark matter is a thermal relic, which in the early universe was in chemical and kinetic equilibrium with the thermal bath of SM particles. When the annihilation rate to the visible sector is greatly enhanced by the resonance, we expect that the product of mediator couplings to DM and the SM particles has to be suppressed to comply with the relic abundance constraints. On the other hand the scatterings of DM particles on the thermal bath states occur only in the t-channel (fig. \ref{feyngraphs}), where a resonance does not emerge, so it remains suppressed. Consequently temperature of the kinetic decoupling $\tkd$ can be much higher than in a typical WIMP scenario ($\tkd\sim\mev$)~\cite{Chen:2001jz}. 

If dark matter decouples kinetically, when it is non-relativistic and its thermal distribution is maintained by self-scatterings\footnote{The resonance may also enhance DM self-interaction, so most of the parameter space considered here corresponds to strongly self-interacting DM.}, then the DM temperature $\tdm$ evolves according to $\tdm\propto R^{-2}$, where $R$ is the cosmic scale factor, contrary to the radiation-dominated SM thermal bath, for which $T\propto R^{-1}$. For the velocity-dependent DM interactions it leads to the modification of the relic density \cite{Dent:2009bv,Bringmann:2006mu,Bringmann:2009vf,vandenAarssen:2012ag}.
We will show that in the case of the BW enhancement, describing kinetic decoupling as instantaneous is not precise~\cite{Bi:2011qm}. When thermally averaged cross-section is temperature dependent, it can affect the DM velocity distribution and consequently also its temperature.  Therefore in order to determine the DM abundance and temperature evolution one has to use a set of coupled  differential Boltzmann equations \cite{Bringmann:2006mu,Bringmann:2009vf,vandenAarssen:2012ag}\footnote{As we expect the temperatures of kinetic and chemical decoupling to be close, we do not neglect $Y_{EQ}$ in (\ref{bey}).}

\bea
\frac{dY}{dx}&=&-\frac{1-\frac{x}{3}\frac{g'_{*s}}{g_{*s}}}{Hx}s\left(Y^2 \sigv_{x_{DM}} -Y_{EQ}^2 \sigv_{x}\right)\non \\
\frac{dy}{dx}&=&-\frac{1-\frac{x}{3}\frac{g'_{*s}}{g_{*s}}}{Hx}\left\{ 2\mdm c(T)(y-y_{EQ}) + \right. \label{bey}\\
 && \left. -\frac{sy}{Y}\left[Y^2\left(\sigv_{x_{DM}}-\sigv_2|_{x_{DM}}\right) - Y^2_{EQ} \left(\sigv_{x}-\frac{y_{EQ}}{y}\sigv_2|_x \right) \right] \right\}\non
\eea

where  $x\equiv\frac{\mdm}{T}$ and $x_{DM}\equiv\frac{\mdm}{T_{DM}}$ and $\mpl=1.22\times 10^{19} \gev$. The yield $Y=n/s$ is expressed by dark matter number density $n$ and entropy density $s=g_{*S}(T)\frac{2\pi^2}{45}T^3$. The equilibrium distribution is $Y_{EQ}(x)=\frac{45}{2\pi^4}\sqrt{\frac{\pi}{8}}\frac{g_i}{g_*}x^{3/2}e^{-x}$, where $g_i$ is the number of DM degrees of freedom. The~Hubble parameter in the radiation-dominated universe is $H=\frac{4\pi^3}{45 \mpl^2}g_* T^4$ and the temperature parameter $y$ is defined as
\beq
y\equiv\frac{\mdm\tdm}{s^{2/3}}\equiv \frac{g_i}{Y s^{5/3}}\int \frac{d^3p}{2\pi^3}\mathbf{p}^2 f(\mathbf{p}) \lsp \text{and} \lsp y_{EQ}\equiv\frac{\mdm T}{s^{2/3}}.
\eeq 
This definition is general, but $T_{DM}$ corresponds to the dark matter temperature only if $f(p)$ is a thermal distribution.
The scattering rate $c(T)$ is given by \cite{Bringmann:2009vf} 
\beq
c(T)=\frac{1}{12(2\pi)^3\mdm^4 T}\sum_f\int dk k^5\omega^{-1} g^\pm(1\mp g^\pm) |\mathcal{M}_f|^2_{t=0; s=\mdm^2+2\mdm\omega+M^2_{f}},
\eeq
where $\mathcal{M}_f$ is the matrix element for the scattering of DM on a given thermal bath state $f$ with momentum $(\omega,k)$ and equilibrium distribution function $g^\pm=(e^{\omega/T}\pm 1)^{-1}$. Finally the averaged cross section $\sigv_2$ reads
\beq
\sigv_{2}|_x = \frac{x^{3/2}}{4\sqrt{\pi}}\int_0^\infty d\vr\sigma\vr\left(1+\frac{1}{6}\vr^2 x\right)\vr^2 e^{-\vr^2 x/4}. 
\eeq

To give a specific example of the interplay between DM density and temperature, we will discuss the evolution of a dark sector within a vector dark matter model (VDM), details of which are presented in the appendix~\ref{model}. This is a Higgs portal scenario, where the second scalar with non-zero VEV is charged under an additional symmetry $U_x(1)$. The extra abelian vector gauge boson $Z'$, which acquires mass due to the Higgs mechanism, can be a stable DM, if its kinetic mixing with the hypercharge gauge boson is forbidden by a $Z_2$ symmetry.
The VDM model contains two scalar mass eigenstates $h_1$ and $h_2$, where the first one is the SM-like Higgs boson and the second will serve as a resonant mediator. The couplings of the scalars are given by
\beq
\begin{split}
\mathcal{L}\supset &\frac{h_1 \cos \alpha + h_2 \sin \alpha}{v}\left(2 M_W W_\mu^+ W^{\mu-} + M^2_Z Z_\mu Z^\mu - \sum_f m_f \bar{f}f \right) +\\
&+(-h_1 \sin \alpha + h_2 \cos \alpha)2 g_x \mzp Z'_\mu Z'^\mu
\end{split}
\eeq
where $\alpha$ is the scalar mixing angle and $g_x$ is the $U_x(1)$ gauge coupling. We focus on the  resonance case $M_{h_2}\approx 2M_{Z'}$ in the limit of a small $\alpha\ll 1$ (appendix~\ref{model}).

\begin{figure}[h!]
 \centering
 \includegraphics[width=0.49\textwidth]{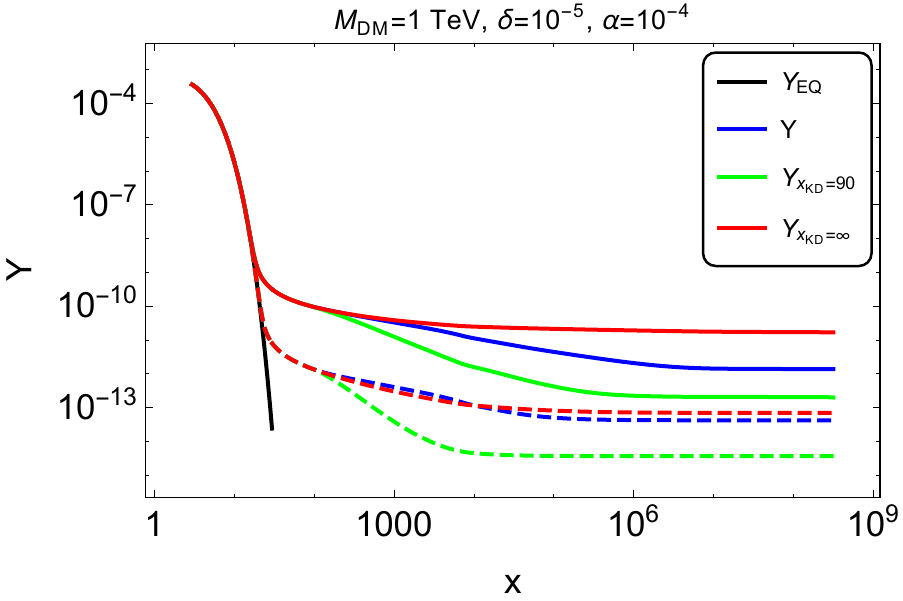}
 \includegraphics[width=0.49\textwidth]{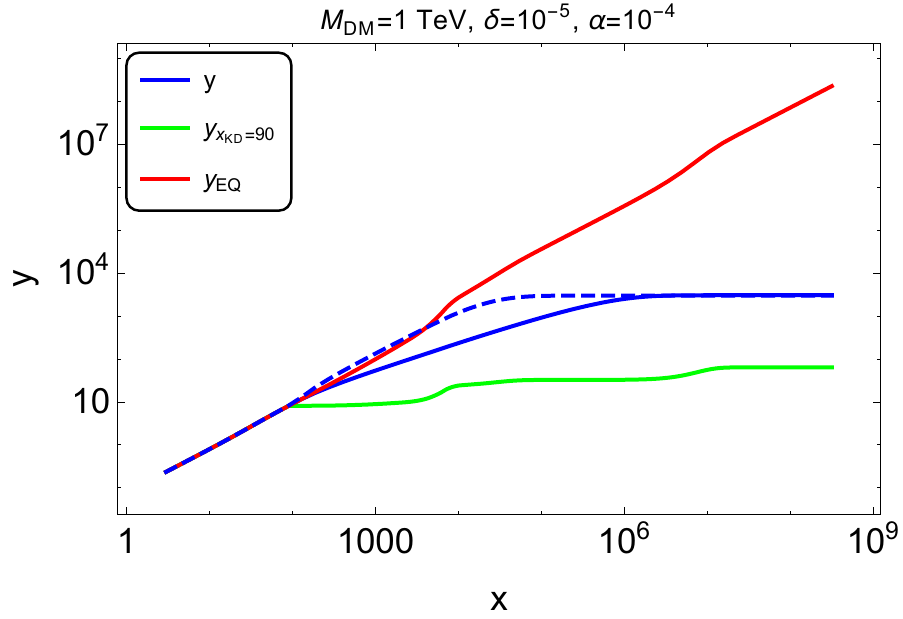}
 \caption{ \label{Yfull} Dark matter yield $Y$ (left panel) and corresponding DM temperatures (right panel)
in different kinetic decoupling scenarios. The blue curves show the solution of the set of equations (\ref{bey}), whereas the green ones refer to the 
instantaneous decoupling at $\xkd=90$. For the red curves dark matter remains in the kinetic equilibrium during its whole evolution. Dashed curves present the corresponding results for the standard Breit-Wigner approximation (with $\gamma\ll\delta$).}
\end{figure}

The figure \ref{Yfull} shows a solution of the full set of equations (\ref{bey}) (blue curve) in comparison to the solution obtained for temperature evolution described by the instantaneous kinetic decoupling (green curve) or assuming no kinetic decoupling (red curve). The value $\xkd=90$ was chosen as the temperature at which $y$ starts to deviate from $y_{EQ}$, which means that temperatures of both sectors are not longer equal. The parameter $y$ settles at constant value, which indicates that dark sector is decoupled ($\tdm\propto T^2$), after long period of evolution. Even though the scattering process is no longer effective the terms in (\ref{bey}), which depend on $Y$ can modify $\tdm$. We note that, as in the case of the Sommerfeld enhancement~\cite{vandenAarssen:2012ag}, this effect can be also substantial for the BW resonance. The cross-section for the annihilation of slower particles is higher than for faster ones, therefore mainly particles with smaller than average kinetic energy are eliminated from the spectrum and the rate at which $\tdm$ decreases is reduced. This interesting behavior ends, when DM annihilation is no longer effective, as can be seen in fig.~\ref{Yfull} by noticing that both $Y$ (left panel) and $y$ (right panel) reach its asymptotic values at the same temperature around~$x\sim 10^6$. 

The non-trivial evolution of dark matter temperature has a strong impact on the final relic abundance. The resonant annihilation cross-section is larger for lower temperatures, therefore $Y$ decreases faster, when $y$ grows at smaller rate. The corresponding solutions obtained for the standard BW approximation with $\gamma\ll\delta$ are depicted in fig.~\ref{Yfull} by the dashed lines. The DM yield $Y$ is smaller in this case, because the cross section $\sigv$ obtained using the BW approximation is overestimated in the region $x\sim 20$, where chemical decoupling happens. Moreover as the annihilation is more efficient, $\tdm$ remains close to the equilibrium temperature for smaller temperatures, therefore the effect of the kinetic decoupling is highly reduced with respect to the previous case. 

\section{Resonant self-scattering of DM}
\label{resonance}

In this and the next sections we will discuss to what extend findings presented in secs.~\ref{res_ann}-\ref{kin_dec} are relevant for the possibility of enhanced DM self-interaction with the s-channel resonance.
Given the dark matter mass $\mdm$ in the GeV range, the cross section $\sigma_{\rm self}$ needs to be of the order of barns, which is many times larger than its typical value for weak interactions. 

The self-interaction cross-section at relative velocity $v_0$  can be written in the vicinity of the resonance as
\beq
\left.\frac{\sigma_{\rm self}}{\mdm}\right\vert_{v=v_0} = \frac{1}{\mdm^3}\frac{8\pi\omega\eta^2_i}{(\delta+v_0^2/4)^2+\gamma^2(v_0)}
\label{resonantcs}
\eeq
In the VDM model, when the self-interaction of dark matter is mediated by the Higgs scalar $h_2$, the width reads
\beq
\gamma(v_0)=\frac{\Gamma_{h_2}(v_0)}{M_{h_2}} = \eta_i \frac{v_0}{2} + \frac{\Gamma_{\rm non-DM}(M_{h_2})}{M_{h_2}},
\label{gam_lab}
\eeq
where $\Gamma_{\rm non-DM}$ includes the width into the SM states and possibly SM-like Higgs. Perturbativity limits the parameter $\eta_i\leq 3/16$ (appendix~\ref{model}). Note that contributions from the SM states (including $h_1$) are suppressed by $\sin^2\alpha$. If we neglect them, and also $\delta$ to get the maximal resonant enhancement, we can obtain an upper bound
\beq
\begin{split}
 \left.\frac{\sigma_{\rm self}}{\mzp}\right\vert_{v=v_0} <  \frac{32\pi\omega}{\mzp^3 v^2_0} < 2.4 \left(\frac{10\km/\s}{v_0}\right)^2 \left(\frac{100\gev}{\mzp}\right)^3 \frac{\cm^2}{g}
 \end{split}
\eeq

In principle the above bound allows for substantial enhancement of the self-interaction cross section at the dwarf galaxy scale ($v_0\sim 10\km/{\rm s}$), if $\zp$ is not too heavy. On the other hand the cross-section is reduced at larger scales in compliance with the cluster bounds. However, when the self-interaction is enhanced by the resonance, the cross-section for annihilation of $\zp\zp$ into SM particles is amplified by the same pole in the propagator. Therefore a tension with indirect searches in dwarf galaxies is expected. We will discuss later these limits in more details. Another remark concerns dark matter direct detection limits. As it has been noticed above, large resonant enhancement requires very small mixing angle $\alpha$, therefore the $Z^\prime$-nucleon scattering cross-section is strongly (by $\sin^2\alpha$) suppressed and experimental constraints are easily satisfied.

We checked also the possibility to obtain large self-interaction with an exchange of the SM-like Higgs boson.
Adopting $2\mzp=\mone$ one finds the upper limit for the cross-section
\begin{equation}
 \left.\frac{\sigma_{\rm self}}{\mzp}\right\vert_{v=v_0} <   \frac{g_x^4}{16\pi\mone}\frac{\sin^4\alpha}{[(1-\sin^2\alpha)\Gamma_{SM}+\eta_i^2\sin^2\alpha\mone v_0/2]^2} < 1.1 \frac{\cm^2}{g},
\end{equation}
where $\Gamma_{SM}$ is the SM width of the Higgs boson.
Using $v_0 = 10 $~km/s, $g_x<\sqrt{2\pi}$ ($\eta_i<3/16$) and $\sin\alpha<0.36$ (the ATLAS and CMS combined limit on $HVV$ coupling \cite{Khachatryan:2016vau}), we found that $\ss$ is in the ballpark of self-interacting DM scenarios. However, in this case self-interaction requires larger values of $\alpha$ and the bounds from indirect searches are even more stringent.

\section{Bounds on the parameter space with resonance enhancement.}
\label{sig_self_fin}
In this section we present numerical results for the annihilation and self-interaction cross-section as a function of $\mdm=\mzp$.
The strategy we adopt is as follows. In order to maximize $\ss$ we choose the maximal coupling in the hidden sector, which is allowed by perturbativity ($\eta=3/16$).
Then we scan over $\mzp$ and $\delta$ and fit $\alpha$ (which suppresses the resonance-SM couplings) to satisfy DM abundance $\odm = 0.1199\pm 0.0022$ at the $3\sigma$ uncertainty \cite{Ade:2015xua}.
In order to estimate $\ss$ we will adopt non-zero velocity $v_0$ which, for dwarf galaxies is of the order of $10$ km/s:

\beq
\frac{\sigma_{\rm self}}{\mzp}= \frac{8\pi\omega}{\mdm^3}\frac{\eta_i^2}{(\delta+v_0^2/4)^2+\gamma^2(v_0)},
\label{ss}
\eeq
where $\gamma^2(v_0)$ is given by (\ref{gam_lab}).   
Then, with $\delta$ chosen, assuming the same DM velocity, we approximate the averaged annihilation cross-section at present, $\sigv_0$, by
\beq
\sigma\vr(v_0)=\frac{16 \pi\omega}{\mzp^2} \frac{\eta_i \eta_f\bar\beta_f}{(\delta+v_0^2/4)^2+
\gamma^2(v_0)}.
\eeq


\begin{figure}[tb]
\includegraphics[width=.43\textwidth]{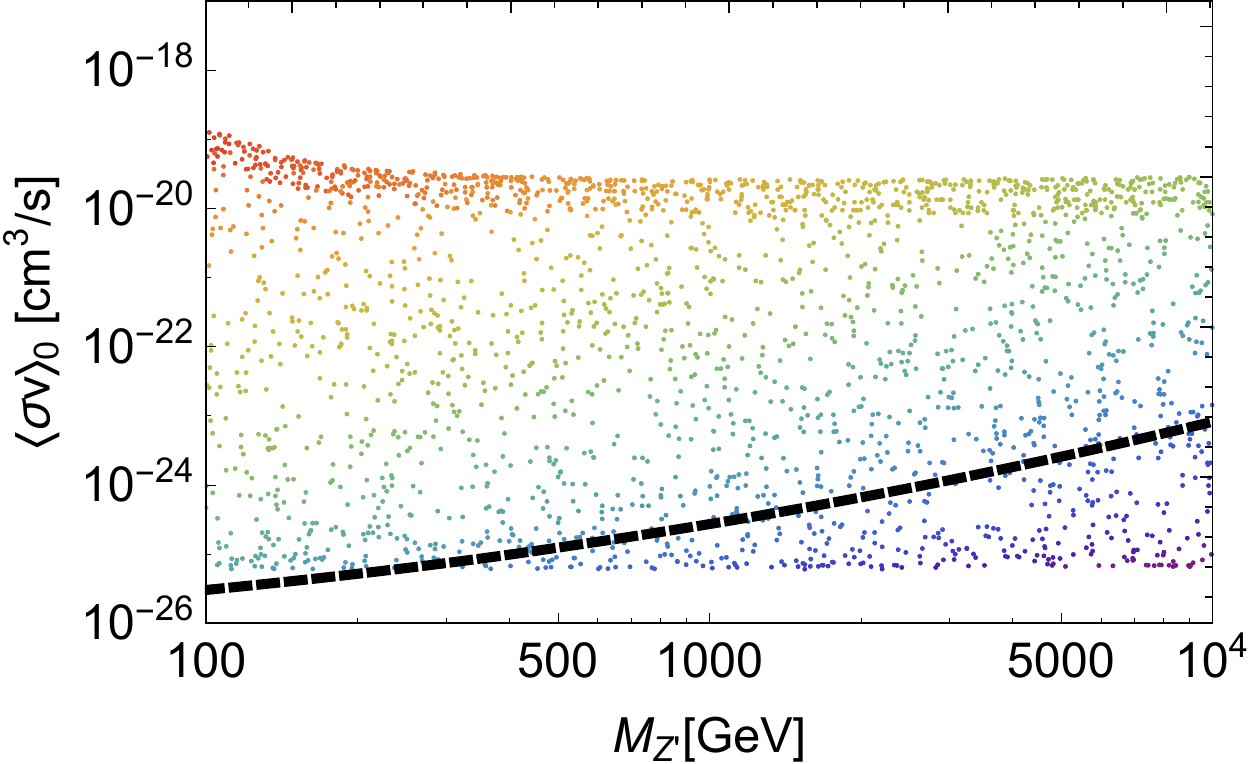}
\includegraphics[width=.43\textwidth]{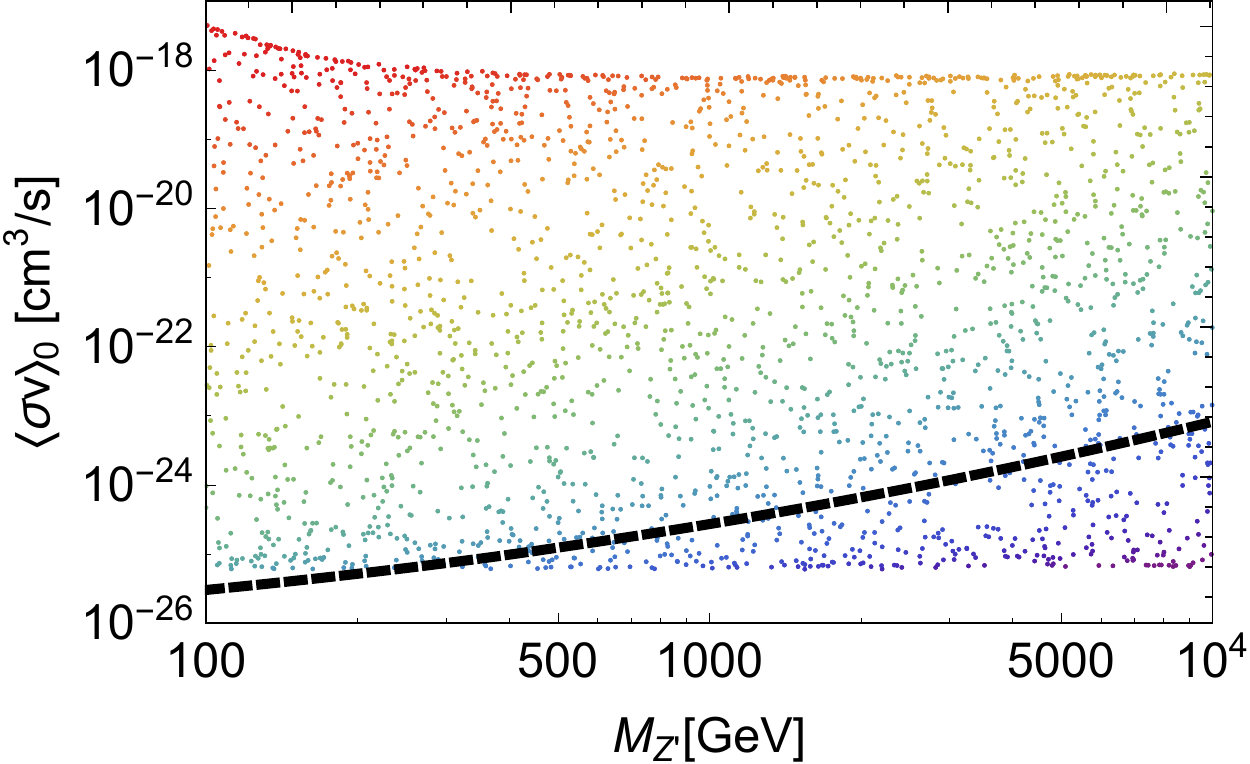}
\includegraphics[width=.12\textwidth]{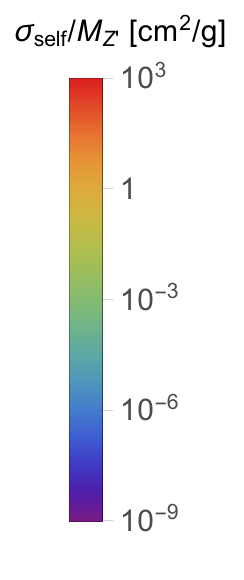}\\
\includegraphics[width=.43\textwidth]{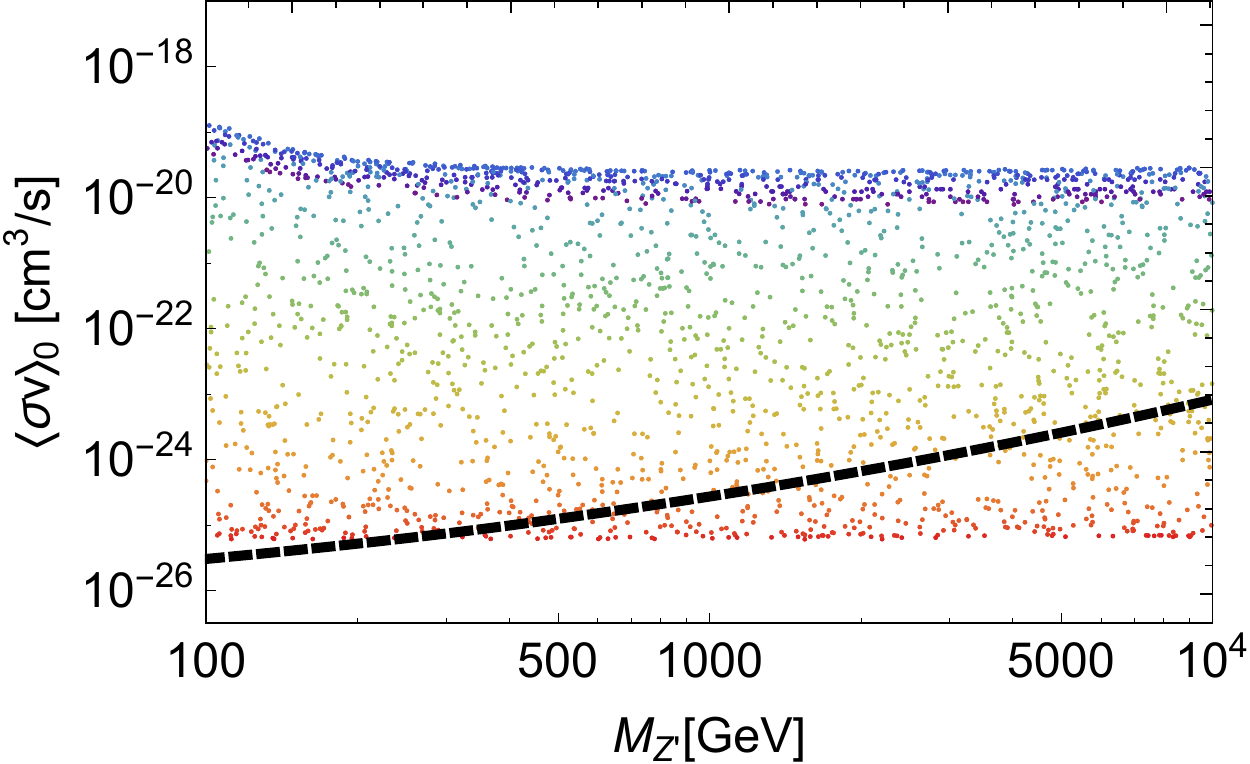}
\includegraphics[width=.43\textwidth]{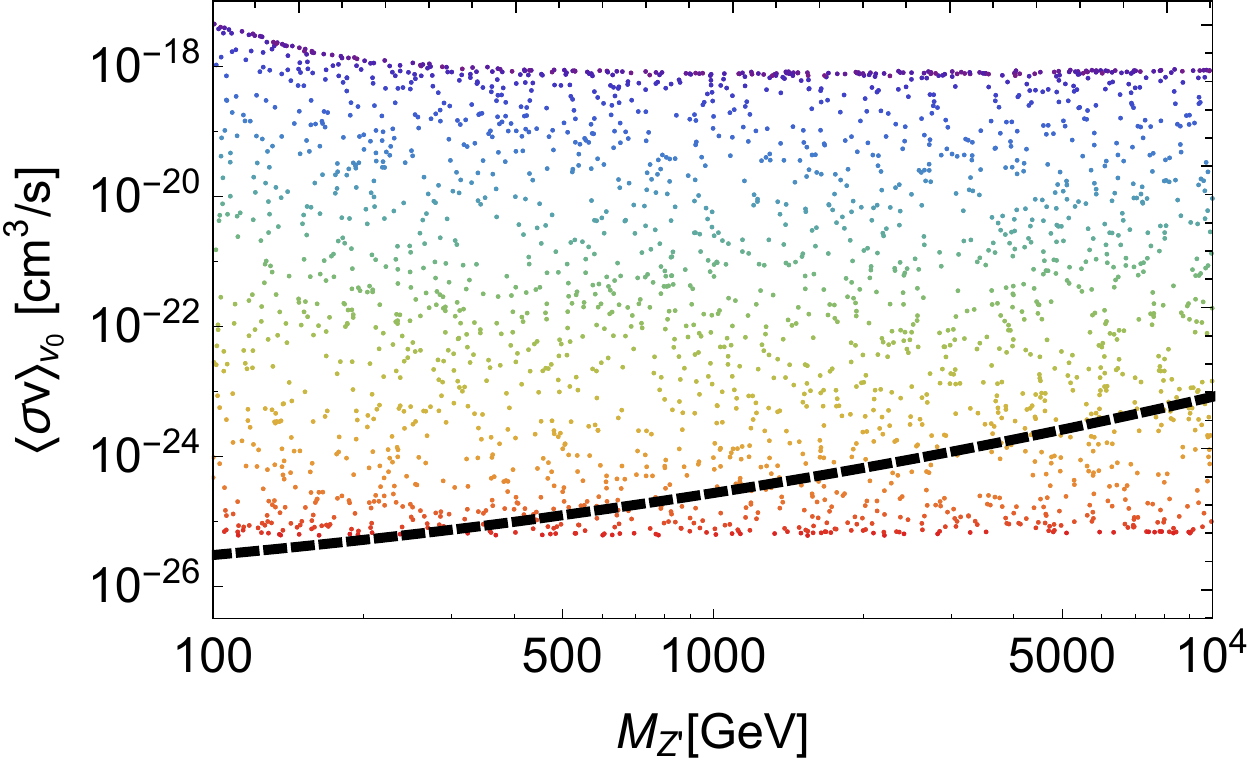}
\hspace{0.23cm}
\includegraphics[width=.075\textwidth]{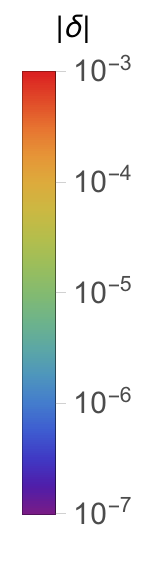}\\
\includegraphics[width=.43\textwidth]{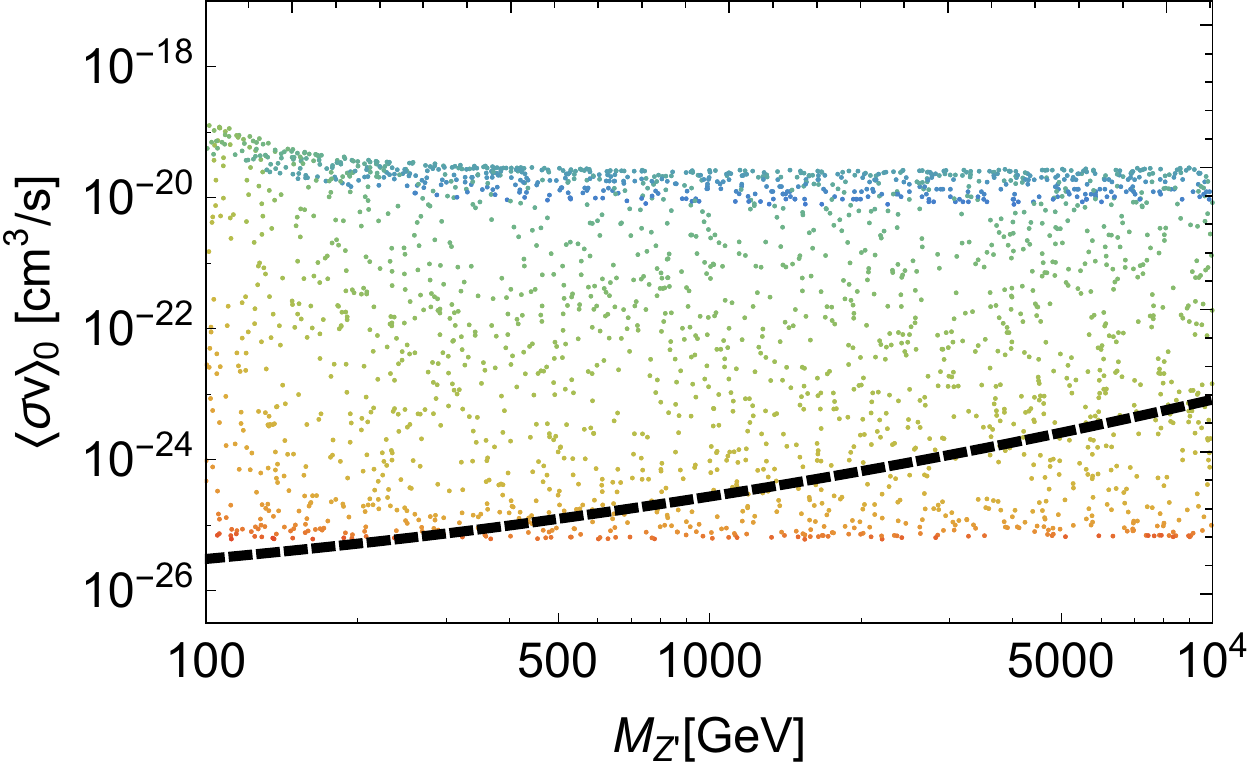}
\includegraphics[width=.43\textwidth]{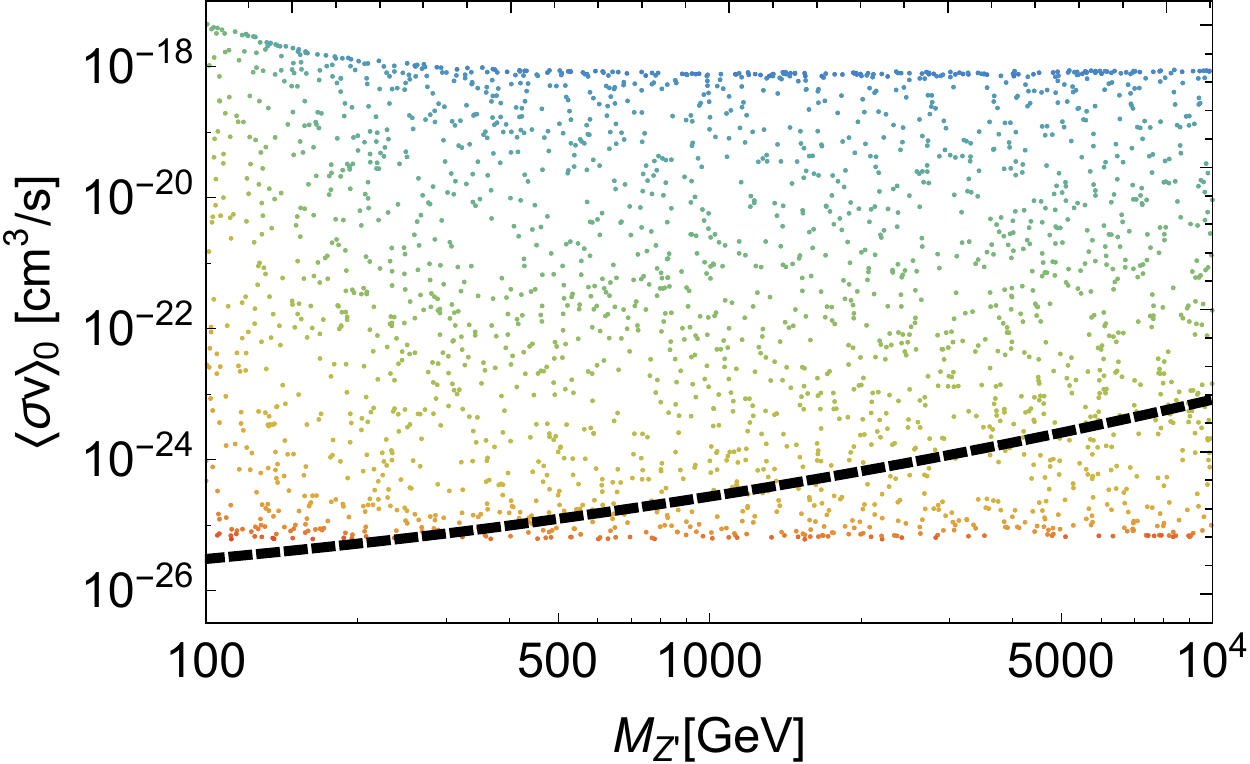}
\hspace{0.23cm}
\includegraphics[width=.075\textwidth]{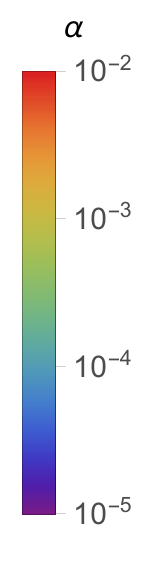}
\caption{\label{scan1}
Results of the scan in the parameter space over $\mzp$ and $\delta<0$. For each point in the plot we fit $\alpha$ to satisfy the relic abundance constraint and then calculate the annihilation $\sigv_{v_0}$ and self-interaction $\ss$ cross-section at the typical velocity $v_0$ equal to $10$~km/s (left panels) and $1$~km/s (right panels). Points are colored with respect to $\sigv_{v_0}$ (top panels),
$\delta$ (middle panels) or $\alpha$ (bottom panels). The maximal value of $\eta_i$ in the VDM model, $\eta=3/16$, was chosen.
}
\end{figure}

Our results are presented in fig. \ref{scan1}. It should be emphasized that all the points shown in the plots correspond to the present abundance of DM calculated taking into account both the early kinetic decoupling via the set of coupled Boltzmann equations (as described in sec.~\ref{kin_dec}) and adopting the velocity dependent resonance ($h_2$) width. For $\sin\alpha \ll 1$ $\Gamma_{\rm non-DM}$ is negligibly small and the total width is dominated by $h_2\to \zp\zp$.

In the upper panels of fig.~\ref{scan1} the coloring corresponds to the value of $\ss$ obtained for a given scan point.
It can be seen that for a given $\mzp$, points corresponding to enhanced $\ss$ are located towards the upper edge of the scanned (in $\delta$) band. 
As one could have easily anticipated large(small) $\sigv_0$ is correlated with large(small) $\ss$, 
so that e.g. it is difficult to reconcile $\ss\simeq 1$~cm$^2$/g and constraints from DM indirect searches. The Fermi-LAT upper limit for the annihilation in the $W^+ W^-$ channel \cite{Ackermann:2015zua} (which is the dominant one in our case) is shown as a dashed line in the figures. The maximal $\ss$ located on the Fermi-LAT curve, turns out to be $\sim 10^{-6}$~cm$^2$/g. In order to see explicitly the tension between large $\ss$ and limited from above $\sigv_0$ it is useful to rewrite the later as follows:

\beq
\sigv_0 = \frac{\Gamma_f}{\eta_i} \left(\frac{\sigs}{\mzp}\right)
\eeq

Our intention was to obtain large $\ss$ while keeping $\sigv_{v_0}$ below the Fermi-LAT upper limit. Naively, one could think that this could be achieved 
suppressing $\Gamma_f$ for all those final states the Fermi-LAT is sensitive to. To achieve that one could adjust small enough coupling between the resonance and the final state $f$, this is however impossible since the relevant parameter, $\sin\alpha$ has already been fixed by the requirement of proper DM abundance. 

To illustrate the velocity dependence we have adopted in figs.~\ref{scan1} two values of the DM velocity: 10~km/s and 1~km/s in the left and right panels, respectively, which are in the range of a typical velocity for dwarf galaxies \cite{Bonnivard:2015xpq}. Larger velocity provides less freedom in enhancing $\ss$. However, note that the lower edge of the allowed band is $v_0$-independent as the minimal $\sigv_0$ is obtained for maximal $\delta$ (as seen from fig.~\ref{scan1}) such that leading velocity-dependent contributions to the denominator $\propto v_0^2$ are negligible. This is an important observation as we are interested in the lowest possible $\sigv_0$. 
Therefore the lower limit for the DM mass is not sensitive the DM velocity. Also the maximal allowed value of $\ss$ (located along the Fermi-LAT curve) is velocity independent.
On the other hand the upper edge for smaller $v_0$ is substantially higher since it suffers from weaker velocity suppression (as it corresponds to small values of $\delta$).

In the middle panels of fig.~\ref{scan1} we show the same results for $\sigv_{v_0}$, however the coloring is with respect to $\delta$. It is important to notice here that for both velocities the smallest $\sigv_{v_0}$ corresponds to the largest $|\delta|$. It could be noticed that for smaller velocity (the lower panel) the cross-section, when starting at the lower edge,  is growing monotonically with decreasing $|\delta|$ while for the larger velocity (upper panel) the minimum of $\delta$ does not refer to the largest cross-section. It happens, because at $\delta\approx \eta v_0$, $\sigv_{v_0}$ changes its scaling behavior from $1/\delta^2$ to $1/v_0^2$ and for further decreasing $\delta$ the cross section is suppressed by smaller values of $\sin^2\alpha$ as required by relic density.

It is also worth to recall here that the Fermi-LAT limits coming from dwarf galaxies include uncertainties coming from  the density profiles (J-factors). There are also extra uncertainties, which arise due to velocity-dependence of the annihilation cross-section, but their inclusion is beyond the scope of this work \cite{Zhao:2016xie}. 

\begin{figure}[tb]
\includegraphics[width=.49\textwidth]{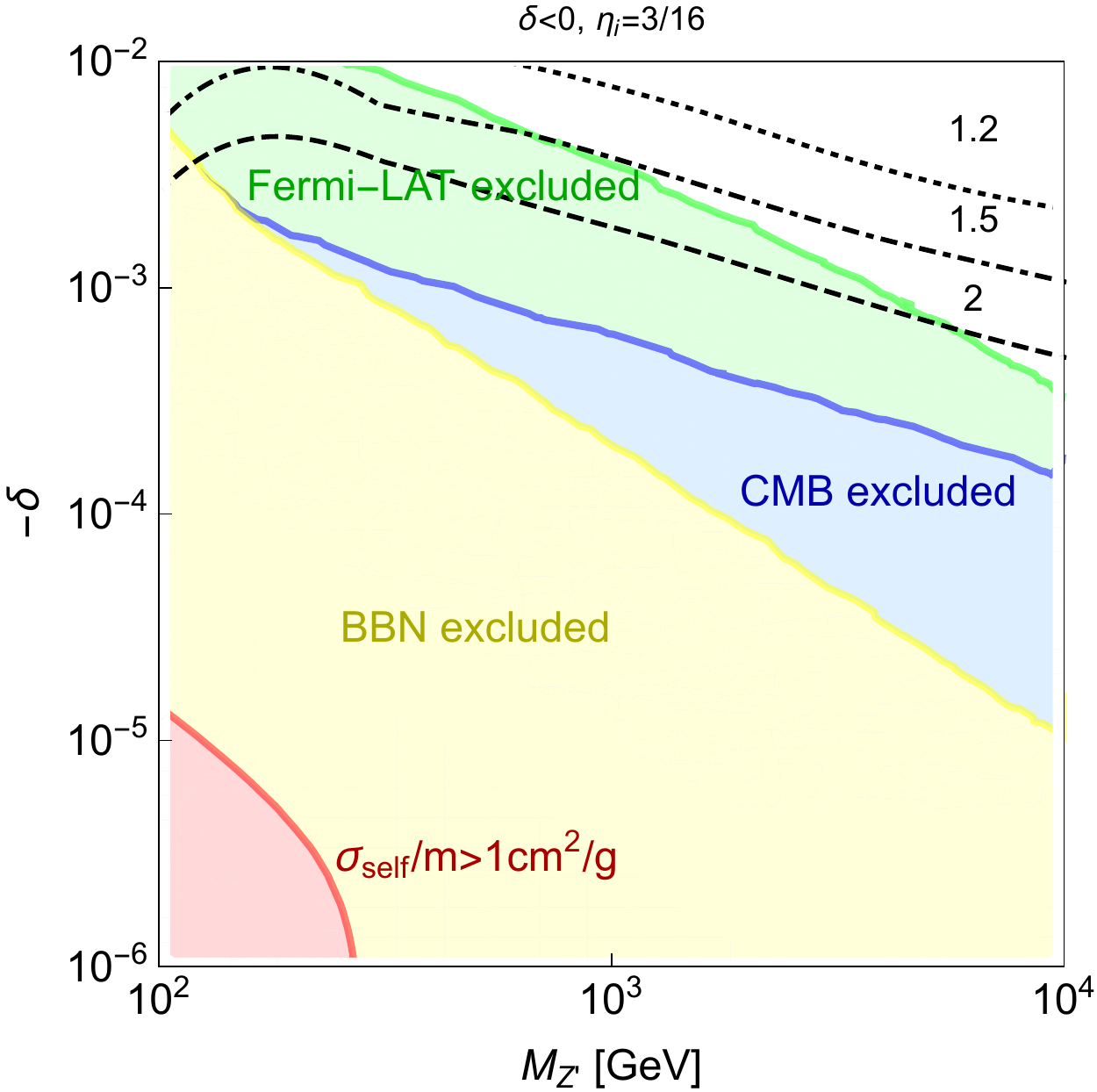}
\includegraphics[width=.49\textwidth]{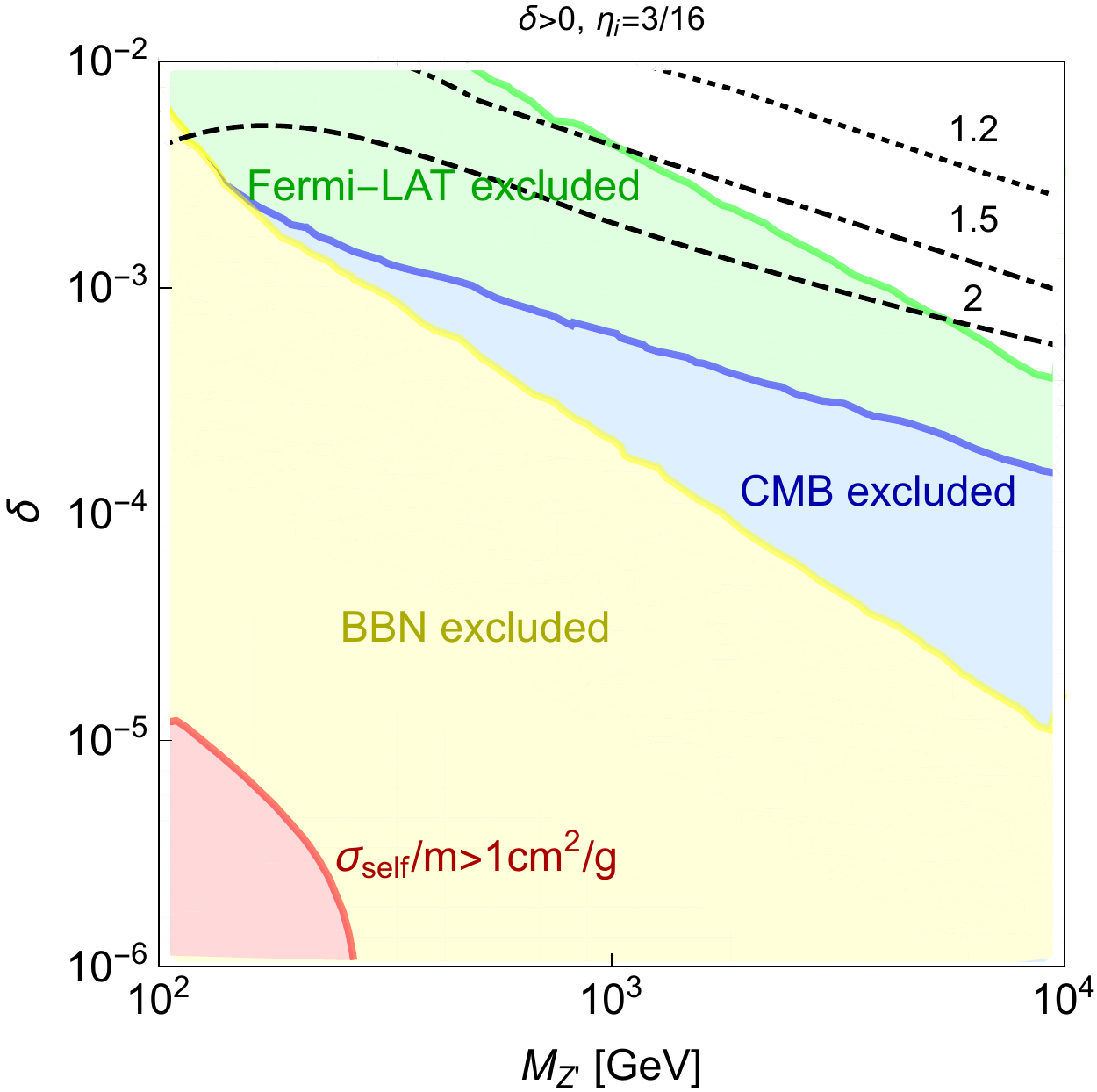}
\caption{\label{constraint}
Constraints coming from Fermi-LAT, CMB, BBN and region with large self-interactions in the plane of $\mzp$ and negative (left panel) or positive (right panel) parameter $\delta$. Below black dotted, dash-dotted or dashed lines relic density without considering kinetic decoupling is larger by factor 1.2, 1.5 or 2 respectively.
}
\end{figure}

In the fig. \ref{constraint} we present various constraints in the parameter space ($\mzp$, $\delta$) of resonantly annihilating dark matter. As before we fit $\alpha$ to satisfy the relic density and obtain the Fermi-LAT bound using  typical velocity $v_0=10$~km/s. The CMB limit on DM annihilation cross-section was taken from \cite{Elor:2015bho} assuming direct annihilation into the SM states. Upper BBN bound comes from the influence of DM annihilation on the deuterium abundance~\cite{Kawasaki:2015yya}. In both cases we assume that annihilation is dominated by the $W^+ W^-$ channel and the cross-section is saturated by small dark matter velocities during BBN or CMB decoupling. Although this maximizes regions excluded by CMB and BBN, we note the Fermi-LAT gives the strongest limit in the whole parameter space and region with large self-interacting cross-section is excluded.

On the other hand the effects of kinetic decoupling on the relic density are substantial also in the allowed region. In case of larger DM masses $\mzp\sim 10$~TeV, the calculation assuming equal temperatures of DM and the visible sector overestimates relic abundance by a factor larger than 2.

\section{Conclusions}
\label{con}
Consequences of resonant contribution to dark-matter annihilation have been discussed in details.
Since dark matter particles are usually non-relativistic therefore in the vicinity of the resonance 
a new threshold for its decays is implicitly present and therefore the standard Breit-Wigner parametrization of the resonance propagator must be modified by momentum dependent contributions to the width. 
Corrections induced by the momentum-dependence were discussed and it has been shown that in the case of non-suppressed coupling between the resonance and the dark matter, the threshold effects are important and can not be neglected. 

The resonance enhancement of the annihilation implies a reduced coupling between resonance and SM and/or resonance and DM. Therefore if one wants to preserve an option of large $\ss$ then the coupling of the resonance to SM must be suppressed  and therefore the early kinetic decoupling of the dark sector has to be considered. Two possible scenarios; a sharp DM-SM temperature splitting and a smooth decoupling were analyzed. It the later case coupled Boltzmann equations had to be solved. It turned out that the early decoupling is relevant and must be properly included.

The above conclusions are generic and should be verified in a given model of DM whenever the resonant enhancement is present.
Here physics in the vicinity of a resonance was illustrated within $U(1)$ vector-boson dark matter model, which, besides the extra (dark) gauge boson, contains a second neutral Higgs boson $h_2$.   
 
The possibility of explaining strong dark-matter self-interaction by an s-channel $h_2$ resonance enhancement was pointed out and analyzed. It has been shown that it is possible to reach moderately enhanced $\ss$ and properly reproduce measured dark matter abundance employing  the second (non-SM like) Higgs boson resonance enhancement. 
However it is not possible to reach the strongly self-interacting DM regime with $\ss = 1\cm^2 / g$.

\section*{Acknowledgments}
The authors thank Torsten Bringmann and Andrzej Hryczuk for discussions and kind information 
concerning their work on a similar subject, which will be completed very soon. 
The authors are also grateful to Apostolos Pilaftsis for inspiring discussions on unitarity and gauge dependence.
This work is partially supported by the National Science Centre (Poland) research project, 
decision DEC-2014/13/B/ST2/03969 and DEC-2014/15/B/ST2/00108.

\appendix
\section{A model of dark gauged \boldmath{$U(1)$} sector}
\label{model}
In this appendix, we are going to describe a model of vector dark matter (VDM) \cite{Hambye:2008bq,Lebedev:2011iq,Farzan:2012hh,Baek:2012se,Baek:2014jga,Duch:2015jta} which is an extension of the SM by an additional $\uone$ gauge symmetry factor together with a complex scalar field $S$.  The vev of the scalar $S$ generates mass for the vector field via the standard Higgs mechanism. The quantum numbers of the scalar field are 
\beq
S  =   (0,{\bf 1},{\bf 1},1) \ \  \text{under}  \ \  U(1)_Y\times SU(2)_L \times SU(3)_c \times \uone.
\eeq
The SM fields are not charged under the extra gauge group.  To ensure stability of the new vector boson
we assume a $\z2$ symmetry to forbid $U(1)$-kinetic mixing between $\uone$ and  $U(1)_Y$. 
The scalar field $S$ and the extra gauge boson $A_\mu$ transform under $\z2$ as follows
\beq
A^{\mu}_X \rightarrow -A^{\mu}_X \ , \ S\rightarrow S^*,  \ \text{where} \  S=\phi e^{i\sigma}, \ \ {\rm so} \ \ \phi\rightarrow \phi,  \ \  \sigma\rightarrow -\sigma.
\eeq
Other fields are neutral under the $\z2$. 

The vector bosons masses are given by:
\beq
M_W=\inv2 g v, \ \ \ \ M_Z = \frac{1}{2}\sqrt{g^2+g'^2} v \ \ \ \text{and} \ \ \   \mzp = \gx \vx,
\eeq
where $g$, $g'$ and $g_x$ are the $SU(2)$, $U(1)$ and $U(1)_X$ gauge coupling constants, respectively, while $v$ and $\vx$ are $H$ and $S$ vacuum expectation values (vev's): $(\langle H \rangle,\langle S \rangle)=\frac{1}{\sqrt{2}}(v,\vx)$.
The scalar potential for this theory is given by
\beq
V= -\mu^2_H|H|^2 +\lambda_H |H|^4 -\mu^2_S|S|^2 +\lambda_S |S|^4 +\kappa |S|^2|H|^2 .
\eeq

The positivity of the potential requires the following constraints that we impose hereafter:
\beq
\lambda_H > 0, \ \ \lambda_S >0, \ \ \kappa > -2 \sqrt{\lambda_H \lambda_S}.
\label{positivity}
\eeq
Both vevs are non-zero and give rise to the masses of the SM fields and dark vector boson,
if $\mu_{H,S}^2<0$ $\kappa^2<4\lambda_H\lambda_S$, then
\beq
v^2=\frac{4 \lambda_S \mu^2_H - 2\kappa \mu^2_S }{4\lambda_H\lambda_S-\kappa^2},\ \ \vx^2=\frac{4 \lambda_H \mu^2_S - 2\kappa \mu^2_H }{4\lambda_H\lambda_S-\kappa^2}.
\label{min_con}
\eeq
The scalar fields shall be expanded around corresponding vev's as follows
\beq
S=\frac{1}{\sqrt{2}}(\vx+ \phi_S +i\sigma_S)  \ \ , \ \  H^0= \frac{1}{\sqrt{2}}(v + \phi_H+ i\sigma_H)   \ \  \text{where} \ \  H=\binom{H^+}{H^0}.
\eeq
The mass squared matrix $\mathcal{M}^2$ for the fluctuations $ \left(\phi_H, \phi_S\right)$ and their eigenvalues $M^2_\pm$ read
\begin{equation} 
\mathcal{M}^2 = \left( 
\begin{array}{cc}
2 \lambda_H v^2  & \kappa v \vx \\ 
 \kappa v \vx &2 \lambda_S v^2_x 
\end{array} 
\right) \ \ , \ \ 
M^2_\pm=\lambda_H v^2 +\lambda_S \vx^2\pm \sqrt{\lambda_S^2 \vx^4 - 2\lambda_H\lambda_S v^2 \vx^2 + \lambda_H^2 v^4 +\kappa^2 v^2 \vx^4}.
\label{massmatrix}
 \end{equation} 
The matrix $\mathcal{M}^2$ could be diagonalized by an orthogonal rotation $R$, such that 
$\mathcal{M}_{\text{diag}}^2= R^{-1} \mathcal{M}^2 R$. The convention adopted for the ordering of the
eigenvalues and for mixing angle $\alpha$ is the following  
\begin{equation} 
\mathcal{M}_{\text{diag}}^2 = \left( 
\begin{array}{cc}
\mone^2  &0 \\ 
0 &\mtwo^2
\end{array} 
\right)   ,
 \ \   \ 
R = \left( 
\begin{array}{cc}
\cos \alpha   & -\sin \alpha \\ 
\sin \alpha &\cos \alpha
\end{array} 
\right), 
 \ \   \
 \left( 
\begin{array}{c}
h_1\\ 
h_2
\end{array} 
\right) = R^{-1}
\left(
\begin{array}{c}
\phi_H\\ 
\phi_S
\end{array} 
\right),
\end{equation} 
where $\mone=\msm\gev$ is the mass of the observed $125\gev$ Higgs particle. Then we obtain
\beq
\tan 2 \alpha = \frac{\kappa v v_x}{\lambda_h v^2 - \lambda^2_s v^2_x} \approx \frac{2\kappa v v_x}{\mone^2-\mtwo^2},
\eeq
where the approximation can be used in the weak coupling limit ($\kappa\ll 1$), which we discus in this paper. In this limit scalar masses are given by $\mone^2=2\lambda_h v^2$, $\mtwo^2=2\lambda_s v^2_x$. Trading $v_x$ with $\mzp/g_x$, we are left with four independent parameter ($\mtwo$, $g_x$, $\mzp$, $\alpha$).

A decay width $h_2\rightarrow \zp\zp$ is given by
\beq
\Gamma_{h_2\rightarrow \zp\zp} = \frac{(2g_x\mzp)^2}{4\mtwo}\left(3-\frac{\mtwo^2}{\mzp^2}+\frac{\mtwo^4}{4\mzp^4}\right)\frac{1}{8\pi}\sqrt{1-\frac{4\mzp^2}{\mtwo^2}}
\eeq
If we limit the quartic couplings by perturbativity to $\lambda_h$, $\lambda_s<4\pi$, then the resonance conditions ($2\mtwo\approx 2\mzp$) leads to the bound $g_x<\sqrt{2\pi}$. Consequently, the parameter $\eta_i$ within the VDM model is limited to
\beq
\eta_i = \frac{\Gamma_{h_2\rightarrow \zp\zp}}{\mtwo\beta_{\zp}} \lsim \frac{3}{16}.
\eeq
\bibliography{biblio} 
\bibliographystyle{JHEP}
\end{document}